\documentclass[twocolumn,showpacs,aps,amsmath,amssymb,superscriptaddress]{revtex4}
\usepackage{graphicx}
\begin{document}
\title{Shot Noise in Magnetic Tunnel Junctions: Evidence for Sequential Tunneling}
\author{R. Guerrero}
\author{F. G. Aliev}
\affiliation{Departamento de Fisica de la Materia Condensada, C-III,
Universidad Autonoma de Madrid, 28049, Madrid, Spain}
\author{Y. Tserkovnyak}
\affiliation{Lyman Laboratory of Physics, Harvard University,
Cambridge, MA 02138 and
            Department of Physics and Astronomy, University of California, Los Angeles, California 90095, USA }
\author{T. S. Santos}
\author{J. S. Moodera}
\affiliation{Francis Bitter Magnet Laboratory, Massachusetts Institute of Technology, Cambridge, MA 02139, USA}
\date{\today}
\begin{abstract}
We report the experimental observation of sub-Poissonian shot noise in single magnetic tunnel junctions,
indicating the importance of tunneling via impurity levels inside the tunnel barrier. For junctions with weak
zero-bias anomaly in conductance, the Fano factor (normalized shot noise) depends on the magnetic configuration
being enhanced for antiparallel alignment of the ferromagnetic electrodes. We propose a model of sequential
tunneling through nonmagnetic and paramagnetic impurity levels inside the tunnel barrier to qualitatively
explain the observations.
\end{abstract}

\maketitle

The discovery of the giant magnetoresistance \cite{Fert88} followed by the observation of a large tunneling
magnetoresistance (TMR) at room temperature in magnetic tunnel junctions (MTJ's) \cite{Moodera95} have boosted
interest in spin-dependent electron transport in magnetic nanostructures, especially in the spin-dependent
tunneling phenomena \cite{RevMoodera99,RevTsymbal03}. During recent years, there has been a growing interest in
controlling the TMR and also the statistics of tunneling events in MTJ's by nanostructuring of the insulating
barrier \cite{Barnas98}. A variety of new electron-correlation mechanisms have been proposed, typically based on
transport through double MTJ's with either an open or Coulomb-blockaded quantum dot (QD) contacted by
ferromagnetic electrodes. Electric shot noise (SN) is a powerful tool for studying correlations of tunneling
processes in nanostructures beyond the capabilities of dc measurements \cite{BB00}. The growing list of
theoretically investigated topics regarding spin-dependent shot noise includes the noise asymmetry between
parallel (P) and antiparallel (AP) ferromagnetic (FM) alignment \cite{Bulka99} as well as continuous variation
of the SN over the relative angle between FM electrodes \cite{Tserkovnyak01}, SN through an artificial (QD)
Kondo impurity \cite{Lopez03} contacted by magnetic leads, shot-noise enhancement by dynamic spin blockade in
tunneling through a small QD \cite{Cottet04}, and shot noise for spin-polarized and entangled electrons with
spin-orbit interaction in the leads \cite{Egues02}. The scope of experimental efforts
\cite{Nowak99,Nowak04,Nowak04II} has however so far been much more limited and inconclusive with regard to the
nature of tunneling electron correlations even in the conceptually simplest spintronic devices, viz. MTJ's, as
manifested by shot-noise measurements.

Current fluctuations due to discreteness of electron charge flowing through the structure out of equilibrium,
which provide the shot noise, contain information not accessible by time-independent conductance. Sensitivity to
quantum statistics, interference, and interactions between electrons passing through the device has made SN an
effective tool for investigating quantum transport in meso- and nano-structures \cite{BB00}. In the absence of
any correlations, Poissonian shot noise is practically frequency independent at low frequencies with the noise
power given by $S=2eI$, in terms of the average current $I$. The Fano factor $F=S/2eI$ representing normalized
shot noise is in general lowered below 1 for noninteracting electrons due to fermionic statistics.
Electron-electron interactions can either further suppress or enhance the Fano factor (even beyond the
Poissonian value).

Despite the theoretical excitement about perspectives of using the shot noise for investigation of
spin-polarized electrons, behavior of the SN even in simple nonstructured MTJ's remains unclear.
Jiang~\textit{et al.} \cite{Nowak04} reported an observation of the ``full" SN (i.e., $F\sim1$) in MTJ's with AP
alignment of electrodes. Later the same group \cite{Nowak04II} measured a strong suppression (down to
$F\approx0.45$) of the SN in magnetic tunnel junctions, which was not understood. Our Letter reports the first
systematic investigation of the tunneling statistics in a magnetic tunneling device by measuring shot noise in
Co(80~\AA )$\mid$Al$_{2}$O$_{3}$(14~\AA )$\mid$Py(100~\AA) MTJ's with and without Cr doping of the insulating
barrier. We demonstrate a decrease of the Fano factor and its dependence on the alignment of the ferromagnetic
electrodes for certain barrier conditions.
%Version 24/10/06
%Experimental observations suggest sequential tunneling through intrabarrier impurity levels.  Cr doping and
%magnetic-configuration control by an applied magnetic field allow us to engineer statistics of spin-polarized
%hopping via the tunnel barrier, opening new venues for studying physics of spin-dependent transport and quantum
%information.

Details of sample preparation have been published previously \cite{Jansen00}. For Cr-doped samples, the tunnel
barriers were deposited in two steps. After deposition of the underlying Co electrode, a first tunnel barrier
was formed by deposition and subsequent oxidation of 7-9~\AA\ of Al. Subsequently, sub-monolayer amounts of Cr
were deposited on the Al$_2$O$_3$ surface, followed by a second Al layer deposition (5-7~\AA) and oxidation,
resulting in a ``$\delta$-doped" Al$_2$O$_3$$\mid$Cr$\mid$Al$_2$O$_3$ tunnel barrier. The noise measurements use
a setup described in Ref.~\cite{Guerrero05}, which employs the cross-correlation method. This technique removes
uncorrelated noise from the amplifiers and the noise of the leads
%CUTs Oct06
%Computer control of the current through the
%sample permits measurements of the noise and dynamic tunneling
%resistance as a function of bias current at fixed magnetic field, in
%this way
and takes into account nonlinearity of the dynamic resistance while converting the obtained voltage noise into
current noise. Out of 13 samples investigated, the shot noise was measured for 11 MTJ's: 5 without and 6 with
$\delta$-layer of Cr in the middle of the barrier, ranging between 0.2 and 1.2~\AA\ in thickness.
%Cuts Oct06
    %The current flowing through the sample is converted into the voltage, which is analyzed either by a dc voltmeter
    %or by a lock-in amplifier to measure the bias voltage and the dynamic conductance, respectively. The voltage
    %from the MTJ's is measured by two identical homemade dc-coupled ultra-low--noise amplifiers placed at the top of
    %the cryostat. The pre-amplified signals are further amplified by additional low-noise amplifiers (Stanford
    %Research SR560). A spectrum analyzer SR780 calculates the cross-correlation spectrum in general containing
    %thermal, shot, and $1/f$ contributions to the noise. The influence of the capacitance of the line (about 400~pF) and of
    %the MTJ's (about 10~nF) as well as their resistance ($10-600$~k$\Omega$) are taken into account during noise
    %analysis without using fitting parameters.
    \begin{figure}
    \includegraphics[width=\linewidth,clip=]{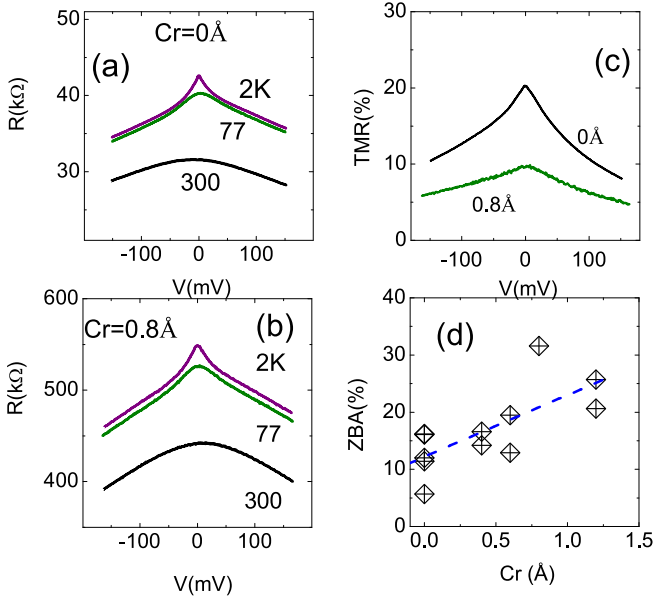}
    \caption{\label{fig1}(Color online) Typical dynamic resistance
    obtained in P state for the Cr-free (a) and Cr-doped (b) junctions with
    0.8~\AA\ $\delta$-layer measured at 300, 77, and 2~K. (c) Reduction of the TMR with applied
    voltage for Cr-free and Cr-doped MTJ's at
    $T=2$~K. (d) Dependence of the ZBA(\%) $=100\times [R(0~{\rm m}V)-R(100~{\rm m}V)]/R(100~{\rm m}V)$ determined for the P alignment
    on Cr (at 2~K).}
    \end{figure}

Figure~\ref{fig1} shows typical electron transport characteristics of the studied MTJ's. The dynamic tunneling
resistance vs bias $V$ [Figs.~\ref{fig1}(a),(b)] measured at three temperatures for P alignment proves
pinhole-free MTJ's \cite{RevMoodera99}. For all MTJ's studied, an asymmetric parabolic conductance background
\cite{brinkman} plus a zero-bias anomaly (ZBA) below $T\sim77$~K, appeared in the resistance of the junction
($R_{J}$) [Figs.~\ref{fig1}(a),(b)].
%change to editor III
    %The general view of the ZBA is that finite temperature and bias allow for some
    %inelastic-scattering processes coupled to tunneling events, opening additional transport channels.
Presently, there exists several possible explanations of the ZBA's in MTJ's
\cite{ZangLevyZBA,JAP98}, which consider magnon- or phonon-assisted tunneling or two-step
tunneling through impurities inside the tunnel barrier which are also coupled to some additional degrees of
freedom.
%Such processes require hot electrons, leading to the low-bias/temperature suppression of the
%conductance.
Simultaneous ZBA and SN measurements on our samples suggest the ZBA is provided by sequential
tunneling through impurities accompanied with spin flips.
    \begin{figure}
    \includegraphics[width=\linewidth,clip=]{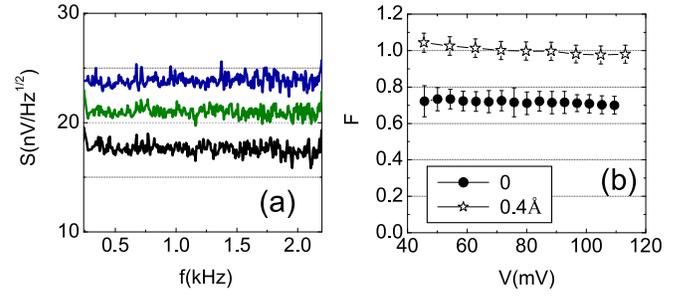}
    \caption{\label{fig2} (Color online) (a) Typical voltage noise
    measured for a Cr-free MTJ at $T=2$~K
    with the applied currents (from bottom to up) of 3.4, 5 and 6~$\mu$A.
    (b) Voltage dependence of $F$ on bias, for the Cr-free (filled) and 0.4~\AA\ Cr-doped
    (open stars), also measured at $T=2$~K. The error bars show
    standard deviations.}
    \end{figure}
%Doping of the barrier with Cr generally suppresses both the TMR and
%the conductivity, but the relations between these parameters and the
%nominal Cr concentration are not strictly monotonic [see, e.g.,
%Fig.~\ref{fig1}(d)].

Doping of the barrier with Cr enhances the normalized ZBA, although this trend presents rather large dispersion
[Fig.~\ref{fig1}(d)]. Conductivity and TMR are generally suppressed when Cr thickness is increased, but the
relations between these parameter and the nominal Cr concentration are not strictly monotonic. We have found,
however, that the changes in the TMR are correlated with those of the tunneling resistance (see below). This can
be understood as follows: As the barrier width and the resistance increase, the relative role of two-step
tunneling increases, which generally reduces the TMR. The TMR is also monotonically reduced with the applied
voltage both for the Cr-free and Cr-doped MTJ's [see Fig.~\ref{fig1}(c)], in accordance with the previous
reports \cite{ZangLevyZBA}.
%The monotonic suppression of the
%TMR and TR with bias indicates we do not have Kondo impurities in the strong-coupling regime (at least down to
%the lowest temperature $T=2$~K), where tunneling resistance decreases at $T, V\to0$ due to the Kondo resonance \cite{KondoIV}.

    \begin{figure}[b]
    \includegraphics[width=\linewidth,clip=]{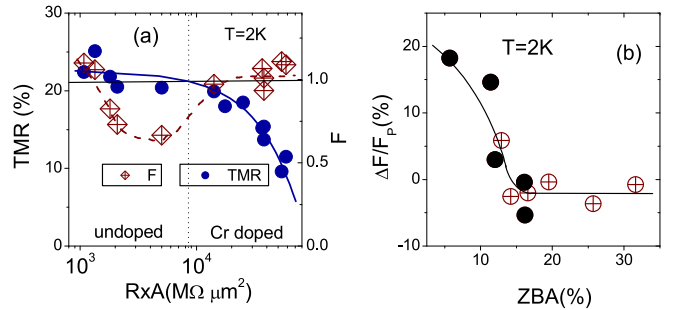}
    \caption{\label{fig3} (Color online) (a) Dependence of the TMR and $F$
    (for the P state) on the resistance area product ($R\times A$).
    Solid horizontal line marks $F=1$. Vertical dashed
    line separates the Cr-free and Cr-doped regions. (b) Dependence of the
    relative variation of $F$ with alignment
    $\Delta F/F_{\rm P}(\%)=100\times(F_{\rm AP}-F_{\rm P})/F_{\rm P}$ on the relative strength of the
    ZBA. Solid symbols point the undoped samples. The lines are guides for the eye.}
    \end{figure}

The measured low-frequency noise has a typical form for MTJ's, with the $1/f$ noise dominating at $f<100$~Hz and
the ``white" noise dominating at $f\gtrsim 100$~Hz. Figure~\ref{fig2}(a) shows a typical voltage noise for the
frequency and bias range where the $1/f$ noise does not affect the data and the applied bias ($eV\gg k_{B}T$)
ensures that SN presents the dominant contribution to the total noise. Fig.~\ref{fig2}(b) shows a typical
dependence of $F$ on bias. For most of the undoped MTJ's, the Fano factor was reduced below the Poissonian value
($F<1$), while for the Cr-doped MTJ's $F$ was always close to one.

Figure~\ref{fig3}(a) shows the TMR and the Fano factor for the P alignment as a function of the resistance by
area product ($R\times A$) at $T=2$~K. The Fano factor was averaged over the range $40-120$~mV where it is
nearly bias independent. For the undoped MTJ's in the range where TMR is only weakly reduced with the product
$R\times A$ ($<10^{4}$~M$\Omega\mu $m$^{2}$), we observed a gradual suppression of the Fano factor down to
$F\sim0.65$. Doping of the barrier with Cr further increases the tunneling resistance and restores the
Poissonian SN ($F\sim1$). The suppression of $F$ in a certain tunneling resistance range is not accompanied by
the appearance of random telegraph noise as in Ref.~\cite{Nowak99}, reduced TMR \cite{shotnoise02}, or by
metallic temperature dependence $R(T)$, clearly ruling out pin-holes/hot spots across the barrier.
Figure~\ref{fig3}(b) shows the normalized AP-P $F$ asymmetry as a function of the normalized ZBA for the P
alignment. Surprisingly, we find that $F$ depends on the alignment of the electrodes with $F_{\rm AP}/F_{\rm P}>1$ only in
the MTJ's with a weak zero-bias anomaly and becomes nearly independent of the alignment above some threshold
value of the ZBA. We stress that the observed Fano factor asymmetry reflects only alignment of the FM
electrodes, but not orientation of the magnetic field.
\begin{figure}[b]
\includegraphics[width=0.9\linewidth,clip=]{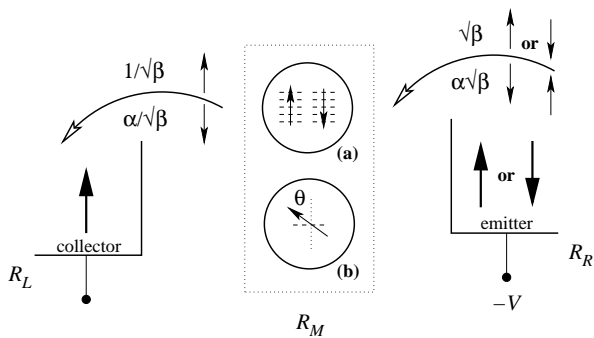}
\caption{Two models: (a) a large normal region $R_M$ such that electrons can be treated as noninteracting,
coupled to biased emitter and collector reservoirs in P or AP configuration, (b) $R_M$ is a single
spin-polarized impurity level that can hold only one electron with spin at a (random) angle $\theta$ with
respect to the collector magnetization.} \label{sc}
\end{figure}
%It was reported previously that for nonmagnetic tunnel junctions
%tunneling through localized states within the barrier could indeed
%account for the measured reduced Fano factor $F<1$ \cite{Iann03}.
%To our best knowledge there have not been

Previous studies of the shot noise in non magnetic TJ's with Al$_2$O$_3$ barrier have observed Poissonian value
$F\simeq 1$ \cite{SNthermometer}. It was reported, however, that for non-magnetic TJ's with SiO$_2$ barrier,
tunneling through localized states within the barrier, could indeed account for the measured reduced $F$
\cite{Iann03}. In the following, we consider two simple models for sequential tunneling via an island inside the
tunnel barrier (see Fig.~\ref{sc}), which capture some qualitative aspects of our measurements. First, consider
tunneling through a normal region ($R_M$) inside the tunnel barrier [Fig.~\ref{sc}(a)]. Neglecting charging
effects, we can simply sum the contributions to the (averaged) current and noise for the two spin species. To
this end, suppose $R_M$ is coupled asymmetrically to the left and right reservoirs ($R_L$ and $R_R$) with the
respective spin-dependent conductances given by
\begin{align}
g_{L\uparrow}=g/\sqrt{\beta}\,\,\,\,\,\,&{\rm and}\,\,\,\,\,\,g_{R\uparrow}=g\sqrt{\beta}\,,\\
g_{L\downarrow}=\alpha g/\sqrt{\beta}\,\,\,\,\,\,&{\rm
and}\,\,\,\,\,\,g_{R\downarrow}=\alpha g\sqrt{\beta}\,
\end{align}
$\beta$ is a dimensionless left-right asymmetry parameter and $\alpha$ characterizes spin polarization. The
charge current at the voltage bias $V$ is given by $I=igV$, parametrized by a dimensionless current $i$ that
depends on $\alpha$ and $\beta$ only. Let us furthermore write the zero-frequency shot noise as $S=2esgV$. The
Fano factor thus becomes  $F=S/(2eI)=s/i$.
\begin{figure}
\includegraphics[width=\linewidth,clip=]{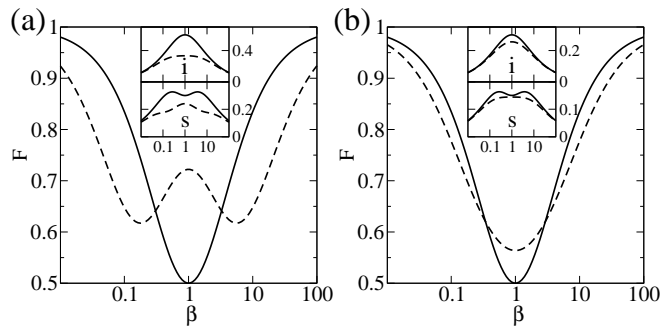}
\caption{Fano factor as a function of the left-right asymmetry parameter $\beta$ setting $\alpha=1/5$ for the two models
sketched in Fig.~\ref{sc}. Solid lines are for the P and dashed for the AP magnetic configurations. Insets show
the dimensionless current $i$ and noise $s=Fi$ defined in the text. Note the logarithmic scale for $\beta$:
Assuming tunnel rates depend exponentially on the barrier thickness, this corresponds to a linear scale for the
$R_M$ position inside the tunnel junction (small $\beta$ corresponding to the proximity to $R_L$ and large
$\beta$ to $R_R$).} \label{theory}
\end{figure}

Let us recall first that, in general, noninteracting spinless electrons in double-barrier structures have the
series conductance and $F$ (for spinless electrons)
%equation 3 in the text
%\begin{equation}
%G=\frac{g_Lg_R}{g_L+g_R}\,\,\,\,\,{\rm
%and}\,\,\,\,\,F=\frac{g_L^2+g_R^2}{(g_L+g_R)^2}\,, \label{GF}
%\end{equation}
\begin{equation}
G=g_{L}g_{R}/(g_{L}+g_{R})\,,\,\,F=(g_{L}^{2}+g_{R}^{2})/(g_{L}+g_{R})^{2}\,, \label{GF}
\end{equation}
which are valid not only for large semiclassical $R_M$ but also for sequential tunneling through a small $R_M$
described by master-equation approach, in which case $g$'s become respective transition rates instead of the
tunnel-barrier conductances \cite{BB00}. Summing corresponding current and noise for the two spin channels in
the P configuration (neglecting correlations between two spin species), one trivially obtains for the current
and $F$ \cite{BB00}
%equation 4 in the text
\begin{equation}
i_{\rm P}=(1+\alpha)\sqrt{\beta}/(1+\beta)\,,\,\,F_{\rm P}=(1+\beta^2)/(1+\beta)^2\,.
\end{equation}
In the AP configuration,
%equation 5 and 6 in the text
\begin{align}
i_{\rm AP}&=\alpha(1+\alpha)(1+\beta)\sqrt{\beta}/[(\alpha+\beta)(1+\alpha\beta)]\,,\\
F_{\rm AP}&=\frac{\alpha^2(1+2\beta-2\beta^2+2\beta^3+\beta^4)}{(\alpha+\beta)^2(1+\alpha\beta)^2}+(\alpha\leftrightarrow\beta)\,,
\end{align}
where $(\alpha\leftrightarrow\beta)$ is the same as the first summand but with $\alpha$ and $\beta$
interchanged. These results are plotted in Fig.~\ref{theory}(a) for $\alpha=1/5$. Note that $F_{\rm AP}-F_{\rm
P}>0$ for $\beta\sim1$, roughly corresponding to the center of the junction for the $R_M$ location, which is the
region contributing the largest current, see Fig.~\ref{theory}(a). The Fano factor asymmetry is reversed closer
to the junction interfaces where the tunneling is asymmetric.

Consider now hopping through a single level that can hold only one extra electron, see Fig.~\ref{sc}(b). If
there is a large exchange-energy splitting along certain direction $\theta$, one could imagine a situation when
only spins polarized along $\theta$ are energetically allowed to tunnel through. We can then calculate the
current and noise using Eqs.~(\ref{GF}) where the rates $g_L$ and $g_R$ now depend on $\theta$ and the relative
magnetic orientation in the leads \cite{Sloncz05}:
%equation 7 and 8in the text
%\begin{align}
%\label{gL}
%g_L&=g_{L\uparrow}\frac{1+\cos\theta}{2}+g_{L\downarrow}\frac{1-\cos\theta}{2}\,,\\
%g_R&=g_{R\uparrow}\frac{1\pm\cos\theta}{2}+g_{R\downarrow}\frac{1\mp\cos\theta}{2}
%\label{gR}
%\end{align}
\begin{align}
\label{gL}
g_L&=g_{L\uparrow}(1+\cos\theta)/2+g_{L\downarrow}(1-\cos\theta)/2\,,\\g_R&=g_{R\uparrow}(1\pm\cos\theta)/2+g_{R\downarrow}(1\mp\cos\theta)/2
\label{gR}
\end{align}
for the P (AP) configuration. Assuming $\theta$ is random, we average the current and noise:
$\langle...\rangle_\theta=(1/2)\int_{-1}^1d(\cos\theta)...$. This results in simple expressions for the P MTJ:
%equation 9 in the text
\begin{equation}
i_{\rm P}=(1+\alpha)\sqrt{\beta}/[2(1+\beta)]\,,\,\,F_{\rm P}=(1+\beta^2)/(1+\beta)^2\,,
\end{equation}
%\begin{equation}
%i_{\rm
%P}=\frac{1}{2}\frac{(1+\alpha)\sqrt{\beta}}{1+\beta}\,\,\,\,\,{\rm
%and}\,\,\,\,\,F_{\rm P}=\frac{1+\beta^2}{(1+\beta)^2}\,,
%\end{equation}
which are the same as just averaging over $\theta=0$ and $\pi$. There is no simple analytic form for the current
and noise in the AP case. We plot the results in Fig.~\ref{theory}(b). Notice that the AP-P asymmetry is
significantly reduced in comparison to Fig.~\ref{theory}(a).

In undoped MTJ's, we measured typically $F_{\rm AP}>F_{\rm P}$ and both are significantly suppressed below 1,
apart from the thinnest tunnel barrier. Both of these findings are consistent with the results in
Fig.~\ref{theory} for tunneling predominantly through impurities in the middle of the barrier.
%As seen from Fig.~\ref{theory}(a), $F$ can be considerably reduced
%below unity even for randomly-distributed impurities within the barrier, as the current is dominated by hopping
%through the center.
Since $F$ for tunneling through uniformly-distributed point-like
localized states is in general $3/4$ (in the absence of hopping
correlations between the two spin species) \cite{Nazarov}, which in
particular applies to both models in Fig.~\ref{sc}, the AP-P
asymmetry would require some structural preference towards tunneling
through the middle of the barrier. The Fano factor is reduced to
$F\sim 3/4$, as shown in Fig.~\ref{fig3}(a), as the tunnel barrier
becomes wider and the role of the two-step tunneling processes
become relatively more important. The tunneling resistance does not
indicate variable-range hopping involving multi-step tunneling,
which was observed for wider tunnel barriers \cite{Xu}. Observation
of Poissonian noise after Cr doping could be due to an offset in Cr
deposited nominally in the center of the junction, which leads to
systematically asymmetric hopping. Finally, the observed correlation
in the AP-P Fano factor asymmetry and the ZBA [Fig.~\ref{fig3}(b)]
suggest that an inelastic spin-flip mechanism in the barrier is
responsible for concurrent reduction of the former and enhancement
of the latter.

In summary, first systematic shot noise measurements in magnetic tunnel junction show
an evidence for sequential tunneling mediated by defects. We demonstrate for the first time that electron
tunneling statistics can be manipulated by an applied magnetic field due to their dependence on the relative
orientation of ferromagnetic electrodes and also by deliberately doping the tunnel barrier with impurities.
Control over the sequential tunneling could find applications in optimizing signal-to-noise ratio in
magnetoelectronic devices and provide a new tool for investigating spin-dependent transport of electrons
injected by ferromagnetic electrodes.
%Finally, presence of sequential tunneling in epitaxial(MgO) barriers through Oxygen vacancies would resolve
%disagreement between theory of coherent tunneling and experiment
%\cite{Tsymbal}

Authors acknowledge P.~LeClair and R.~Villar for critical reading of
the manuscript. The work at UAM was supported in parts by Spanish
MEC (MAT2003-02600, MAT2006-07196, 05-FONE-FP-010-SPINTRA) and CM
(NANOMAGNET). The work at MIT was supported by NSF and KIST-MIT
project grants.


\begin{thebibliography}{}
%\bibitem{Fert88} M.~N.Baibich, J.M.Broto, A.Fert, F.N.Van Dau, F.Petroff, P.Eitenne, G. Creuzet, A. Friederich, and J. Chazelas, Phys. Rev. Lett. {\bf 61}, 2472 (1988).
\bibitem{Fert88} M.~N. Baibich \textit{et al.}, Phys. Rev. Lett. {\bf 61}, 2472 (1988).
\bibitem{Moodera95} J.~S. Moodera \textit{et al.}, Phys. Rev. Lett. {\bf 74}, 3273 (1995);
%\bibitem{Moodera95} J.~S. Moodera, L. R. Kinder, T. M. Wong, and R. Meservey, Phys. Rev. Lett. {\bf 74}, 3273 (1995).
T.~Miyazaki and N.~Tezuka, J. Magn. Magn. Mater. {\bf 139}, L231 (1995).
\bibitem{RevMoodera99} J.~S. Moodera, J.~Nassar, and G.~Mathon, Ann. Rev. Mater. Science {\bf 29}, 381 (1999).
\bibitem{RevTsymbal03} E.~Y. Tsymbal, O.~N. Mryasov, and P.~R LeClair, J. Phys.: Condens. Matter {\bf 15}, R109 (2003);
%\bibitem{Yuasa04} S.~Yuasa, T.Nagahama, A.Fukushima, Y.Suzuki, and K.Ando, Nature Materials {\bf 3}, 868 (2004).
S.~Yuasa \textit{et al.}, Nature Materials {\bf 3}, 868 (2004);
%\bibitem{Parkin04} S.S.~P. Parkin, C.Kaiser, A.Panchula, P.M.Rice, B.Hughes, M.Samant, and S.H.Yang, Nature Materials {\bf 3}, 862 (2004).
S.~S.~P. Parkin \textit{et al.}, \textit{ibid.} {\bf 3}, 862 (2004).
\bibitem{Barnas98} J.~Barna\'{s} and A.~Fert, Phys. Rev. Lett. {\bf 80}, 1058 (1998); S.~Takahashi and S.~Maekawa, \textit{ibid.} {\bf 80}, 1758 (1998).
\bibitem{BB00} Ya.~M. Blanter and M.~B{\"{u}}ttiker, Phys. Rep. {\bf 336}, 1 (2000).
%\bibitem{Russek} N.A.~Stutzke, S.~L.~Burkett, S.~E.~Russek, Appl.Phys.Lett. {\bf 82}, 91 (2003).
%\bibitem{Bulka99} B.R. Bulka, J. Martinek, G. Michalek,  J. Barnas, Phys. Rev. B {\bf 60}, 12246 (1999).
\bibitem{Bulka99} B.~R. Bulka \textit{et al.}, Phys. Rev. B {\bf 60}, 12246 (1999).
\bibitem{Tserkovnyak01} Y.~Tserkovnyak and A.~Brataas, Phys. Rev. B {\bf 64}, 214402 (2001).
\bibitem{Lopez03} R.~Lopez and D.~Sanchez, Phys. Rev. Lett. {\bf 90}, 116602 (2003).
\bibitem{Cottet04} A.~Cottet, W.~Belzig, and C.~Bruder, Phys. Rev. Lett. {\bf 92}, 206801 (2004).
\bibitem{Egues02} J.~C. Egues, G.~Burkard, and D.~Loss, Phys. Rev. Lett. {\bf 89}, 176401 (2002).
%\bibitem{Nowak04} L.~Jiang, E.R. Nowak, P.E. Scott, J. Johnson, J.M. Slaughter, J.J. Sun, R. W. Dave, Phys. Rev. B {\bf 69}, 054407 (2004).
\bibitem{Nowak04} L.~Jiang \textit{et al.}, Phys. Rev. B {\bf 69}, 054407 (2004).
%\bibitem{Nowak04II}L. Jiang, J.F. Skovholt, E.R.Nowak, J.M Slaughter, Proceedings of SPIE, {\bf 5469}, 13 (2004)
\bibitem{Nowak04II}L.~Jiang \textit{et al.}, Proceedings of SPIE {\bf 5469}, 13 (2004)
\bibitem{Nowak99} E.~R. Nowak, M.~B. Weissman, and S.~S.~P. Parkin, Appl. Phys. Lett. {\bf 74}, 600 (1999)
%\bibitem{shotnoise02}  P.K. George, Y. Wu, R.M. White, E. Murdock, M. Tondra, Appl. Phys. Lett. {\bf 80}, 682 (2002)
\bibitem{Jansen00} R.~Jansen and J.~S. Moodera, Phys. Rev. B {\bf 61}, 9047 (2000).
%\bibitem{Guerrero05} R.~Guerrero, F. G. Aliev, R. Villar, J. Hauch, M. Fraune, G. G{\"{u}}ntherodt, K. Rott, H. Br{\"{u}}ckl, and G. Reiss, Appl. Phys. Lett. {\bf 87}, 042501 (2005).
\bibitem{Guerrero05} R.~Guerrero \textit{et al.}, Appl. Phys. Lett. {\bf 87}, 042501 (2005).
\bibitem{brinkman} W.~F. Brinkman, R.~C. Dynes, and J.~M. Rowell, J. Appl. Phys. {\bf 41}, 1915 (1970).
%\bibitem{Guntherodt01} U.~May, K.Samm, H.Kittur, J.Hauch, R.Calarco, U.R{\"{u}}digier, G.G{\"{u}}ntherodt, Appl. Phys. Lett. {\bf 78}, 2026 (2001).
\bibitem{ZangLevyZBA} S.~Zhang \textit{et al.}, Phys. Rev. Lett. {\bf 79}, 3744 (1997);
%\bibitem{MooderaZBA} J.~S. Moodera, J.~Nowak, R.~J.~M. van~de~Veerdonck, Phys. Rev. Lett. {\bf 80}, 2941 (1998).
J.~S. Moodera \textit{et al.}, \textit{ibid.} {\bf 80}, 2941 (1998).
\bibitem{JAP98} J.~Zhang and R.~White, J. Appl. Phys. {\bf 83}, 6512 (1998); L.~Sheng, D.~Y. Xing, and D.~N. Sheng, Phys. Rev. B {\bf 70}, 094416 (2004).
%\bibitem{KondoIV} S.~Bermon, D.~E. Paraskevopoulos, and P.~M. Tedrow, Phys. Rev. B {\bf 17}, 2120 ( 1978); J.~Martinek {\em at al.}, Phys. Rev. Lett. {\bf 91}, 127203 (2003); S.-Y. Bae and S.~X. Wang, IEEE Trans. Mag. {\bf 38}, 2721 (2002).
%\bibitem{Egues94} J.~C. Egues, S.~Hershfield, and J.~W. Wilkins, Phys. Rev. B {\bf 49}, 13517 (1994).
\bibitem{shotnoise02}  P.~K. George \textit{et al.}, Appl. Phys. Lett. {\bf 80}, 682 (2002).
\bibitem{SNthermometer} L.~Spietz \textit{et al.}, Science, {\bf 300}, 5627 (2003).
%\bibitem{Iann03} G.~Iannaccone, F. Crupi, B. Neri, and S. Lombardo, IEEE Trans. El. Dev. {\bf 50}, 1363 (2003).
\bibitem{Iann03} G.~Iannaccone \textit{et al.}, IEEE Trans. El. Dev. {\bf 50}, 1363 (2003).
\bibitem{Sloncz05} J.~C. Slonczewski, Phys. Rev. B {\bf 71}, 024411 (2005).
\bibitem{Nazarov} Y.~V. Nazarov and J.~J.~R. Struben, Phys. Rev. B {\bf 53}, 15466 (1996).
\bibitem{Xu} Y.~Xu, D.~Ephron, and M.~R. Beasley, Phys. Rev. B {\bf 52} 2843 (1995).
%\bibitem{Tsymbal} Private communication.
\end{thebibliography}
\end{document}